\documentclass[traditabstract]{aa}
\usepackage{txfonts}
\usepackage{natbib}
\usepackage{graphicx}
\usepackage{rotating}
\usepackage{aalongtable}
\usepackage{float}
\newcommand{\Msun}{M$_{\odot}$}

\newcommand{\iso}{ISO-ChaI 217}
\newcommand{\para}{Par-Lup 3-4 }
\newcommand{\xsh}{X-Shooter }
\newcommand{\Lsun}{L$_{\odot}$}

\newcommand{\Rsun}{R$_{\odot}$}

\newcommand{\Ha}{H$\alpha$}

\newcommand{\km}{km~s$^{-1}$}

\newcommand{\ISO}{ISO-ChaI 217 }

\begin{document}

\title {Accretion-ejection connection in the young brown dwarf candidate ISO-ChaI 217\thanks{Based on Observations collected with X-shooter at the Very Large Telescope on Cerro Paranal (Chile), operated by the European Southern Observatory (ESO). Program ID: 089.C-0143(A).}}

\author{Whelan, E.T. \inst{1}
  \and 
 Alcal{\'a}, J.M. \inst{2}
 \and
 Bacciotti, F.  \inst{3}
 \and
  Nisini, B. \inst{4}
  \and 
  Bonito, R. \inst{5, 6}
  \and
   Antoniucci, S. \inst{4}
   \and
   Stelzer, B. \inst{6}
  \and
Biazzo, K. \inst{2}
\and
D'Elia, V. \inst{7}
\and
 T.P. Ray. \inst{8}}

\institute{Institut f{\"u}r Astronomie und Astrophysik, Kepler Center for Astro and Particle Physics, Eberhard Karls Universit{\"a}t,  72076 T{\"u}bingen, Germany 
 \and  
 INAF-Osservatorio Astronomico di Capodimonte, via Moiariello, 16, I-80131, Napoli, Italy 
  \and 
  INAF-Osservatorio Astrofisico di Arcetri, Largo E. Fermi 5, I-50125 Firenze, Italy 
  \and 
  INAF-Osservatorio Astronomico di Roma, via Frascati 33, I-00040 Monteporzio Catone, Italy 
  \and  
  Dipartimento di Fisica e Chimica, Universit{\`a} di Palermo, Piazza del Parlamento 1, I-90134 Palermo, Italy 
  \and 
  INAF-Osservatorio Astronomico di Palermo, Piazza del Parlamento 1, I-90134 Palermo, Italy
\and 
ASI – Science Data Center, Via del Politecnico snc, I-00133 Rome, Italy
\and 
Dublin Institute for Advanced Studies, 31 Fitzwilliam Place, Dublin 2, Ireland  }

\titlerunning{X-Shooter Spectroscopy of ISO-ChaI 217} 
\date{}

\abstract  {As the number of observed brown dwarf  outflows is growing it is important to investigate how these outflows compare to the well-studied jets from young stellar objects. A key point of comparison is the relationship between outflow and accretion activity and in particular the ratio between the mass outflow and accretion rates ($\dot{M}_{out}$/$\dot{M}_{acc}$). The brown dwarf candidate ISO-ChaI 217 was discovered by our group, as part of a spectro-astrometric study of brown dwarfs, to be driving an asymmetric outflow with the blue-shifted lobe having a position angle of $\sim$ 20$^{\circ}$. The aim here is to further investigate the properties of \iso, the morphology and kinematics of its outflow, and to better constrain $\dot{M}_{out}$/$\dot{M}_{acc}$. The outflow is spatially resolved in the $ [\ion{S}{ii}]\, \lambda \lambda 6716,6731$ lines and is detected out to $\sim$ 1\farcs6 in the blue-shifted lobe and 1~\arcsec\ in the red-shifted lobe. The asymmetry between the two lobes is confirmed although the velocity asymmetry is less pronounced with respect to our previous study. Using thirteen different accretion tracers we measure log($\dot{M}_{acc}$) [\Msun/yr]= -10.6 $\pm$ 0.4. As it was not possible to measure the effect of extinction on the \iso\ outflow $\dot{M}_{out}$ was derived for a range of values of A$_{v}$, up to a value of A$_{v}$ = 2.5~mag estimated for the source extinction. The logarithm of the
mass outflow ($\dot M_{\rm out}$) was estimated in the range -11.7 to -11.1 for both jets combined. Thus $\dot{M}_{out}$/$\dot{M}_{acc}$ [\Msun/yr] lies below the maximum value predicted by magneto-centrifugal jet launching models. Finally, both model fitting of the Balmer decrements and spectro-astrometric analysis of the \Ha\ line show that the bulk of the H I emission comes from the accretion flow.   }


\maketitle

\section{Introduction}
Jets and outflows are an integral part of the star formation process and models of jet launching and propagation developed for low mass young stellar objects (YSOs) may also 
apply to a diverse range of astrophysical objects, including jets driven by brown dwarfs (BDs) \citep{Frank14, Joergens13}. BDs are now routinely observed in star forming regions and therefore it is important to understand how they form and evolve \citep{Luhman12}. A key way to do this is to investigate their accretion and outflow properties and compare them to YSOs. Since evidence first emerged that BDs launch outflows \citep{Comeron01, Whelan05} the number of BD outflows, both atomic and molecular has grown to $\sim$ 10 \citep{Phan08, Whelan12, Joergens12, Stelzer13, Monin13}. While this is still a statistically small sample, the comparison between BD outflows and those driven by Class II YSOs i.e classical T Tauri stars (CTTSs) has proven to be very interesting \citep{Whelan09b, Whelan11}.  Similarities include the observation of a knotty jet which is a signature of episodic accretion \citep{Whelan12}, the detection of molecular components to known BD optical outflows \citep{Phan08, Monin13}, the presence of both high and low velocity components to the outflows \citep{Whelan09a} and the discovery of an asymmetric BD jet \citep{Whelan09b, Joergens12}. 

\begin{figure*}
\centering
   \includegraphics[width=16cm, trim= 0cm 2cm 0cm 4.5cm, clip=true]{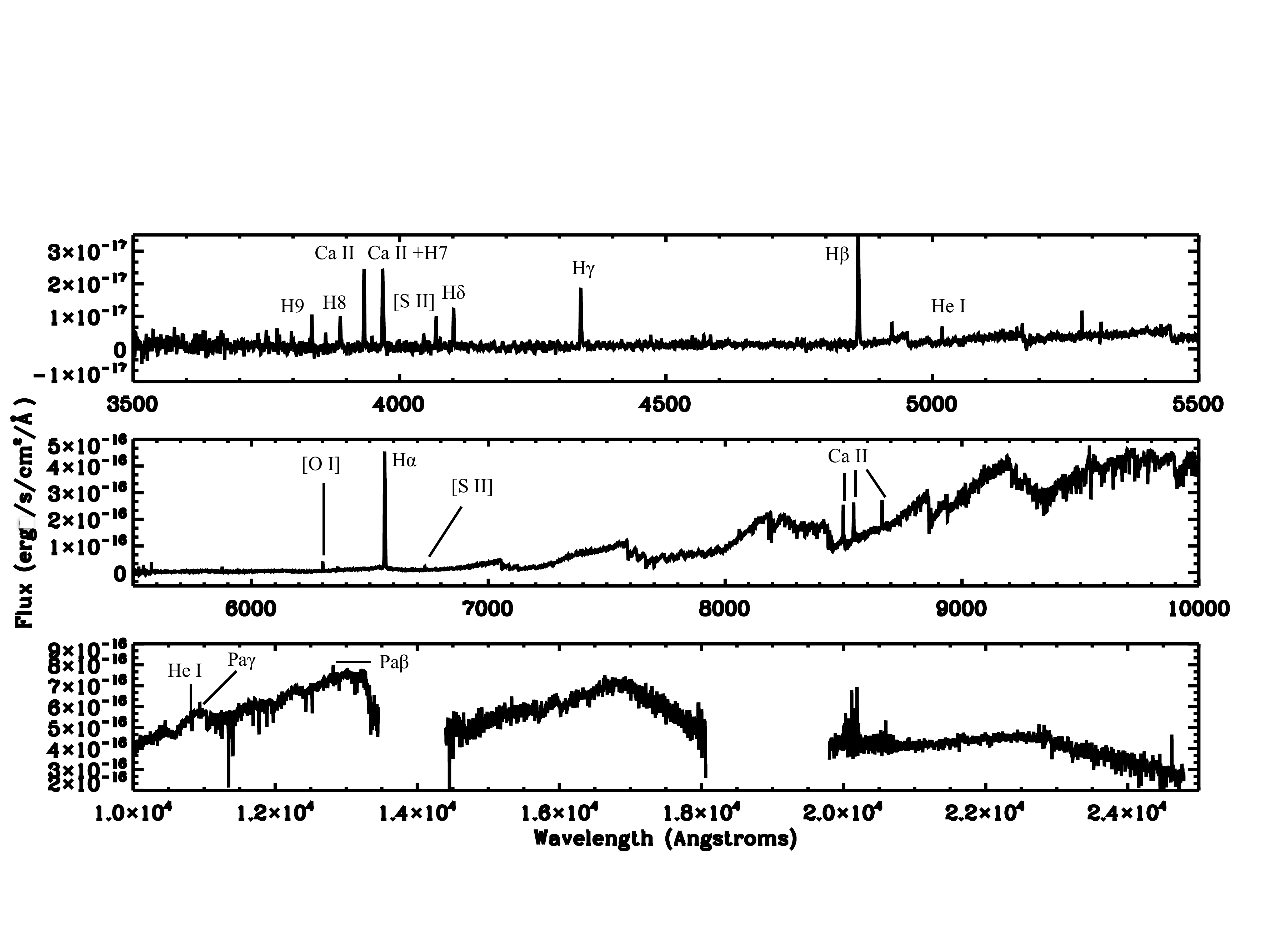}
     \caption{The full \xsh spectrum of ISO-ChaI 217. Key lines are marked and the fluxes of these lines are given in Table 1. Here the spectrum has been smoothed to increase the SNR of key features. The spectra have not been extinction corrected.}
  \label{full}     
\end{figure*}

The high angular resolution observations needed to directly probe the appropriateness of jet launching models developed for YSOs, in the BD mass regime, are not currently possible due to the faintness of the BD outflow emission \citep{Ray07}. However, the ratio of the mass outflow to accretion rate ($\dot{M}_{out}$/$\dot{M}_{acc}$) is constrained by jet launching models and can be investigated in BDs. The first attempts at measuring this ratio in BDs yielded values of $\dot{M}_{out}$/$\dot{M}_{acc}$ which were significantly higher than the $\sim$ 10$\%$ measured for low mass YSOs  and than the predictions made by magneto-centrifugal jet launching models \citep{Whelan09b}. Magneto-centrifugal jet launching models place an upper limit (per jet) of $\sim$ 0.3 on this ratio \citep{Ferreira06, Cabrit09}. However, as discussed in \cite{Whelan09b, Whelan14}, observational biases had a strong influence on these first studies. How this ratio compares in BDs and low mass YSOs could tell us something very important about how BDs form. Therefore, observations specifically designed for measurements of $\dot{M}_{out}$/$\dot{M}_{acc}$, and of a greater number of BDs are needed.
 
\begin{figure}
\centering
   \includegraphics[width=7cm, trim= 2cm 2cm 2cm 4cm]{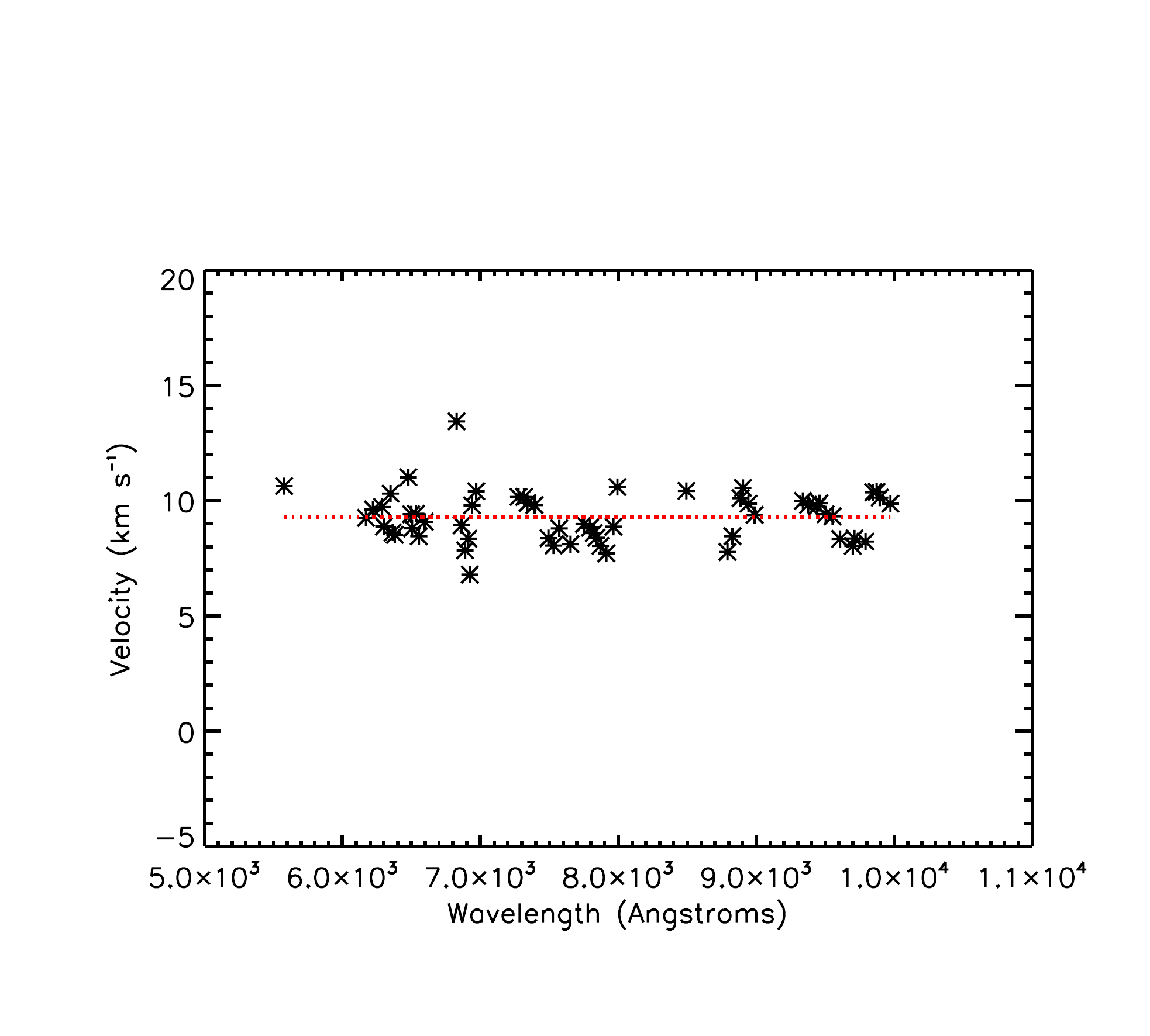}
     \caption{The velocity of various telluric lines across the wavelength range of the VIS arm of \xsh. The difference between the calibration performed by the pipeline and the velocity of the telluric lines is $\sim$ 9 kms$^{-1}$.}
  \label{shift}     
\end{figure}

 \begin{table*}
\centering
\begin{tabular}{rrrrrrr}
\hline \hline 
\multicolumn{2}{c}{\hspace{2cm} Identification}\\
\cline{1-4}
$\lambda_{air}$ (\AA)  &Ion  &Type &E$_{u}$ (cm$^{-1}$)   &Jet &$\lambda_{obs}$ (\AA)  &Flux    
 \\  												
3734.4	&	H I	   &	 13	- 2                                                              &	109029.8    &   	&	3734.3	&	0.6 $\pm$ 0.5
\\												
3750.2	&	H I	   &	 12	- 2                                                   &	108917.1  &     	&	3750.1	&	0.8 $\pm$ 0.5
\\																							
3770.6	&	H I	   &	11 - 2	                                                   &	108772.3	&		&	3770.4	&	0.7  $\pm$ 0.4
\\												
3797.9	&	H I	   &	10 - 2	                                                  &	108582.0	&		&	3797.5	&	1.1 $\pm$ 0.4
\\												
3835.3	&	H I	    &	9 - 2 	                                                           &	108324.7	&		&	3835.4	&	3.0 $\pm$ 0.4
\\																								
3889.0	&	H I	    &	8 - 2 	                                                           &	107965.1	&		&	3888.9	&	2.7 $\pm$ 0.4
\\												
3933.7	&	Ca II 	    &	$^2$P$_{3/2}$-$^2$S$_{1/2}$	            &	25414.4	   &       		&	3933.5	&	5.6 $\pm$ 0.3
\\																							
3968.2$^a$	&	Ca II 	&	$^2$P$_{1/2}$-$^2$S$_{1/2}$              &	25191.5 &	          		&3968.2    &4.5 	$\pm$ 0.3
\\												
3970.1	&	H I	&	 7 - 2	                                                             &	107440.5	& &3969.9				 &2.1 	$\pm$ 0.3	
\\																							
4101.7	&	H I	&	6 - 2	                                                              &	106632.2	  & &			4101.5	&	2.6 $\pm$ 0.3
\\																								
4340.5	&	H I	         &	 5 - 2	                                                        &105291.7 &			&	4340.2	&	4.2 $\pm$ 0.2
\\																																
4861.3	&	H I	&	4 - 2	&	102823.9		&	&	4860.9	&	11.2 $\pm$ 0.15
				
\\ \\
\hline
\\												
5875.9	&	He I 	&	$^3$D$_{1}$-$^3$P$_{0}$	&	186101.7	 &		      &	5876.2	&	2.9 $\pm$ 0.2
\\																								
6300.3	&	[O I]	&	$^1$D$_{1}$-$^3$P$_{2}$	&	15867.9	&Blue		&	6300.3	&	4.25 $\pm$ 0.2
\\	
                   &                &                                                               &                            &Red                       &6300.3                          &4.25  $\pm$ 0.2 
\\ 												
6363.7	&	[O I]	&	$^1$D$_{2}$-$^3$P$_{1}$	&	15867.9	&		&	6363.7	&	2.6 $\pm$ 0.2
\\																																		
6562.8	&	H I	&	3 - 2	                                              &	97492.3	&		&	6563.1	&	195.0 $\pm$ 0.2
\\																							
6583.5	&	[N II]	&	$^1$D$_{2}$-$^3$P$_{2}$	&	15316.2	&Blue		&	6583.3	&	1.2 $\pm$ 0.2
\\																								
6678.2	&	He I 	&	$^1$D$_{2}$-$^1$P$_{1}$	&	186105.1	&		&	6678.8	&	1.5 $\pm$ 0.2
\\												
6716.4	&	[S II]	&	$^2$D$_{5/2}$-$^4$S$_{3/2}$	&	14884.7	&Blue		&	6716.4	&	1.1 $\pm$ 0.2
\\	
                   &                &                                                               &                          &Red                       &6717.7                          &1.25  $\pm$ 0.2  
\\																										
6730.8	&	[S II]	&	$^2$D$_{3/2}$-$^4$S$_{3/2}$	&	14852.9	&Blue		&	6731.7	&	1.8 $\pm$ 0.2
\\	
                   &                &                                                               &                          &Red                       &6732.0                          &2.2  $\pm$ 0.2  
\\																										7065.2       &      He I   &      $^3$S$_{1}$-$^3$P$_{0}$            &	183236.9   &                &     7065.8       &1.0   $\pm$ 0.2
\\    	
7172.0       &      	[Fe II] &	$a^2$G$_{7/2}$-$a^4$F$_{7/2}$  &     16369.4      &                &    7172.4        &0.8  $\pm$ 0.2 
\\	
7432.3        &		[Fe II] &	$b^2$F$_{5/2}$-$a^2$P$_{3/2}$ &31811.82  &               &  7432.9          &1.5  $\pm$ 0.2
\\			
8498.0	&	Ca II 	&	$^2$P$_{3/2}$-$^2$D$_{3/2}$	&	25414.4	&		&	8498.6	&25.8 $\pm$ 0.2
\\												
8542.1	&	Ca II 	&	$^2$P$_{3/2}$-$^2$D$_{5/2}$	&	25414.4	 &		&	8542.5	&	38.4 $\pm$ 0.2					
\\
8662.1	&	Ca II 	&	$^2$P$_{1/2}$-$^2$D$_{3/2}$	&	25191.5	 &		&	8662.5	&	30.2 $\pm$ 0.2					
\\ \\																			
 \hline
 \\  																												
10938.0	&	H I	&	0 - 3 	&		106632.2	 &	&	10937.9	&	33.2 $\pm$ 0.3
\\																					
12818.1	&	H I	&	 5 - 3 	&	105291.7	 &		&	12817.7	&	56.5 $\pm$ 0.5												
\\																				
\hline\hline
\end{tabular}
\caption{Emission Lines identified in the spectrum of \ISO. Fluxes are in units of erg/s/cm$^{2}$ $\times$ 10$^{-17}$ and are not corrected for extinction. \tablefoottext{a}{blended lines.}}
\end{table*}


In \cite{Whelan09b} visible spectra of a sample of BDs at orthogonal slit position angles (PA) was obtained. Spectro-astrometry (SA) was used to confirm that the BDs were driving outflows and the orthogonal data allowed the PAs of the outflows to be constrained. As a follow-up to this discovery study, \iso\ the target with the strongest jet emission and most interesting properties, was observed with X-Shooter. The aim of these observations was spatially resolve the \iso\ outflow by placing the slit along the estimated outflow PA and to refine earlier measurements of $\dot{M}_{out}$/$\dot{M}_{acc}$. Here the results of this \xsh study of \iso, are presented. The full \xsh spectrum is shown in Figure \ref{full} and the fluxes of identified lines are given in Table  1. The fundamental parameters of \iso\ are derived and discussed in Section  3.1 and the morphology and kinematics of the jet are discussed in Section 3.3. $\dot{M}_{acc}$ is calculated from the luminosity of various accretion tracers and $\dot{M}_{out}$ from the luminosity of the $[\ion{S}{ii}]\, \lambda 6731$ line, using the approach of \cite{Whelan14} (see Section 4.1). Finally, the origin of the permitted emission in the spectrum of \iso\ is investigated through the Balmer decrements (Section 4.2). This work is part of a larger study to investigate the accretion-ejection connection in BDs and very low mass stars (VLMSs) \citep{Whelan14, Gi13, Stelzer13}


\section{Target, observations and analysis}

\subsection{Target}
\iso\ (11$^{h}$09$^{m}$52$^{s}$.2, -76$^{\circ}$39$^{\arcmin}$12.$^{\arcsec}$8) is a young very low mass object located in the Chamaeleon I dark cloud (d$\sim$140 $\pm$ 20~pc; Persi et al. 2000). {\bf Its spectral type has been reported as being M6.25 and comparison with models give it a mass of 80~M$_{JUP}$, a radius of 0.64~\Rsun\ and luminosity L$_{bol}$ = 0.023 to 0.028 \citep{Muz05, Luhman07}}. This mass estimate places it at the hydrogen burning mass limit therefore it is reasonable to describe it as a good candidate for being a young BD. Describing it as a BD candidate is also consistent with previous studies of this object. {\bf \cite{Luhman07} assign an age of 5-6~Myr to the northern part of the cloud in which \iso\ lies.} \cite{Muz05} also report log($\dot{M}_{acc}$) = -10 \Msun yr$^{-1}$ for \iso. This estimate of $\dot{M}_{acc}$ is derived by fitting the H$\alpha$ line profile with magnetospheric accretion models. Note that the authors do not give any error estimate for this value of $\dot{M}_{acc}$. \cite{Whelan09b, Whelan14b} measured the PA of the blue-shifted outflow at 20$^{\circ}$ $\pm$ 10$^{\circ}$ and reported its interesting asymmetry. The asymmetry was revealed in the relative brightness of the two lobes of the outflow (the redshifted lobe was brighter), the difference in the radial velocity of the two (the redshifted lobe was faster) and the difference in the electron density (again higher in the red lobe). \cite{Whelan09b} also estimated $\dot{M}_{out}$/$\dot{M}_{acc}$ $>$ 1 for \iso. The results of \cite{Whelan09b} were upheld by a subsequent investigation of the properties of \iso\ by \cite{Joergens12}. Moreover, \cite{Joergens12} investigated the properties of the disk of \iso\ inferring a disk inclination angle of $\sim$ 45$^{\circ}$ and a disk mass of 4 $\times$ 10$^{-6}$~\Msun. \cite{Toro14} report the possible detection of a companion to ISO-ChaI 217 (ISO-ChaI 217 B) with a binary separation of $\sim$ 5~AU and a PA of 238$^{\circ}$ $\pm$ 8$^{\circ}$. The presence of a companion which is interacting with the disk of the jet source could explain the observed asymmetry between the lobes of the jet.

\subsection{Observations}
The \xsh observations presented in this paper were conducted on April 17 2012, on the VLT as part of the INAF \xsh GTO program on star forming regions \citep{alcala11}. The single-node exposure time 
was 850 sec, yielding a nominal exposure time of almost 2 hours after 2 (ABBA) cycles. The average seeing was 0\farcs8 during the observations. The slit was aligned with the jet axis and the slit widths of the UVB, VIS and NIR arms were 1\farcs0, 0\farcs9 and 0\farcs9 respectively.  
This choice of slit widths yielded
spectral resolutions of 5100, 8800 and 5600 for each arm, respectively. The pixel scale is 0\farcs16 for the UVB and VIS arms and 0\farcs21 for the NIR arm. More details of the observations of the GTO program can be found in \cite{alcala11} and \cite{Alcala14}.

\begin{figure}
\centering
   \includegraphics[width=8cm, trim= 1cm 8cm 1cm 8cm]{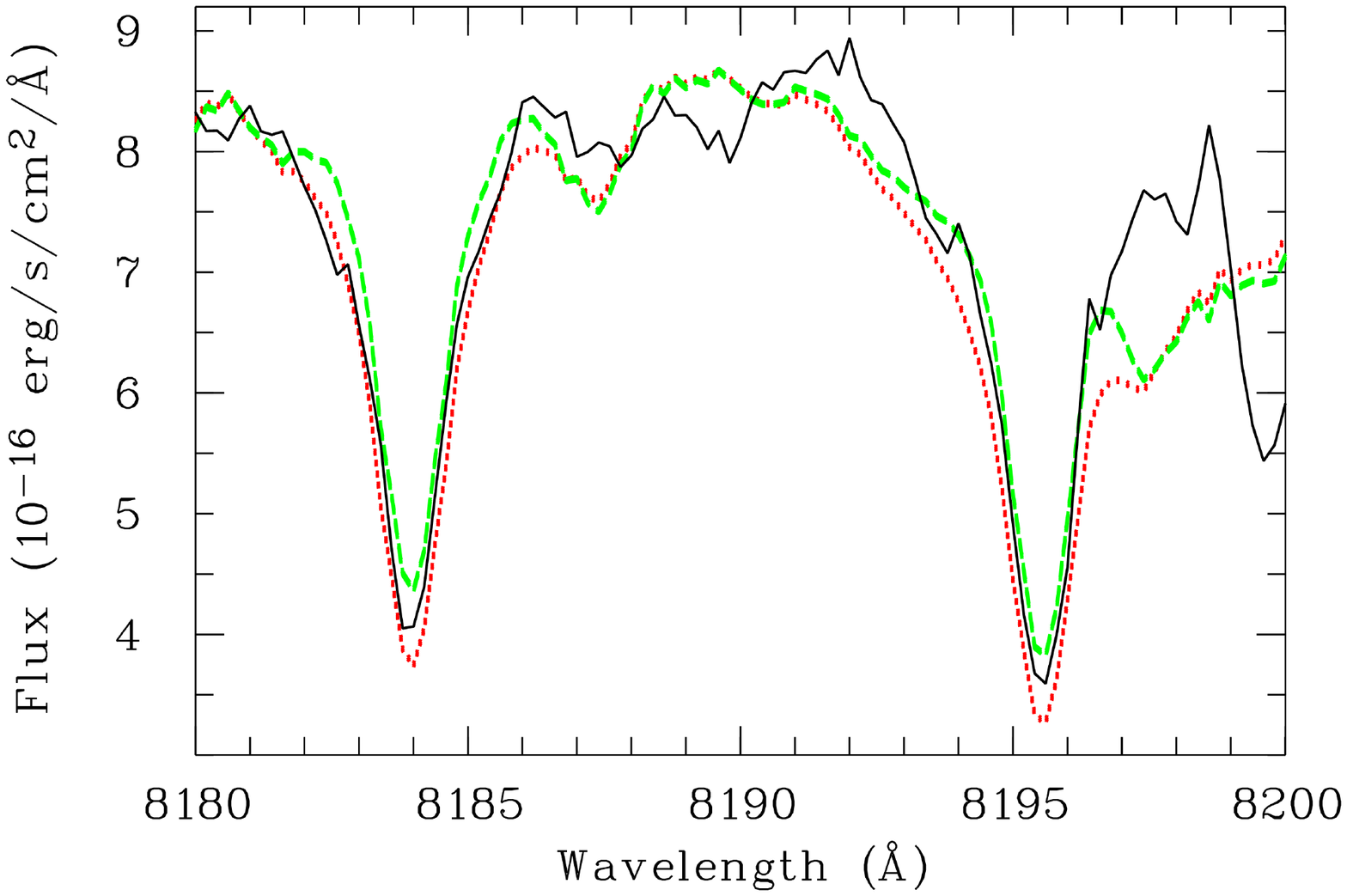}
    \caption{Detail of the X-shooter spectrum of ISO~Cha~I-217 (black solid line) 
     in the wavelength range around the  $\ion{Na}{i}$ $\lambda\lambda$ 8183.3, 8194.8 \AA\ 
     absorption doublet. 
     The spectrum of \iso\ has been corrected for extinction and for telluric absorption lines. 
     The green and the red dotted lines  represent synthetic spectra with 
     $\log{g}$ values of 3.5 and 4.0, respectively, for  $T_{\rm eff}=$2950\,K and 
     rotationally broadened to $v\sin{i}=$20\,km\,s$^{-1}$.}
  \label{logg-vsini}     
\end{figure}

\begin{figure*}
\centering
   \includegraphics[width=16cm, trim= 0cm 2cm 0cm 3cm, clip=true]{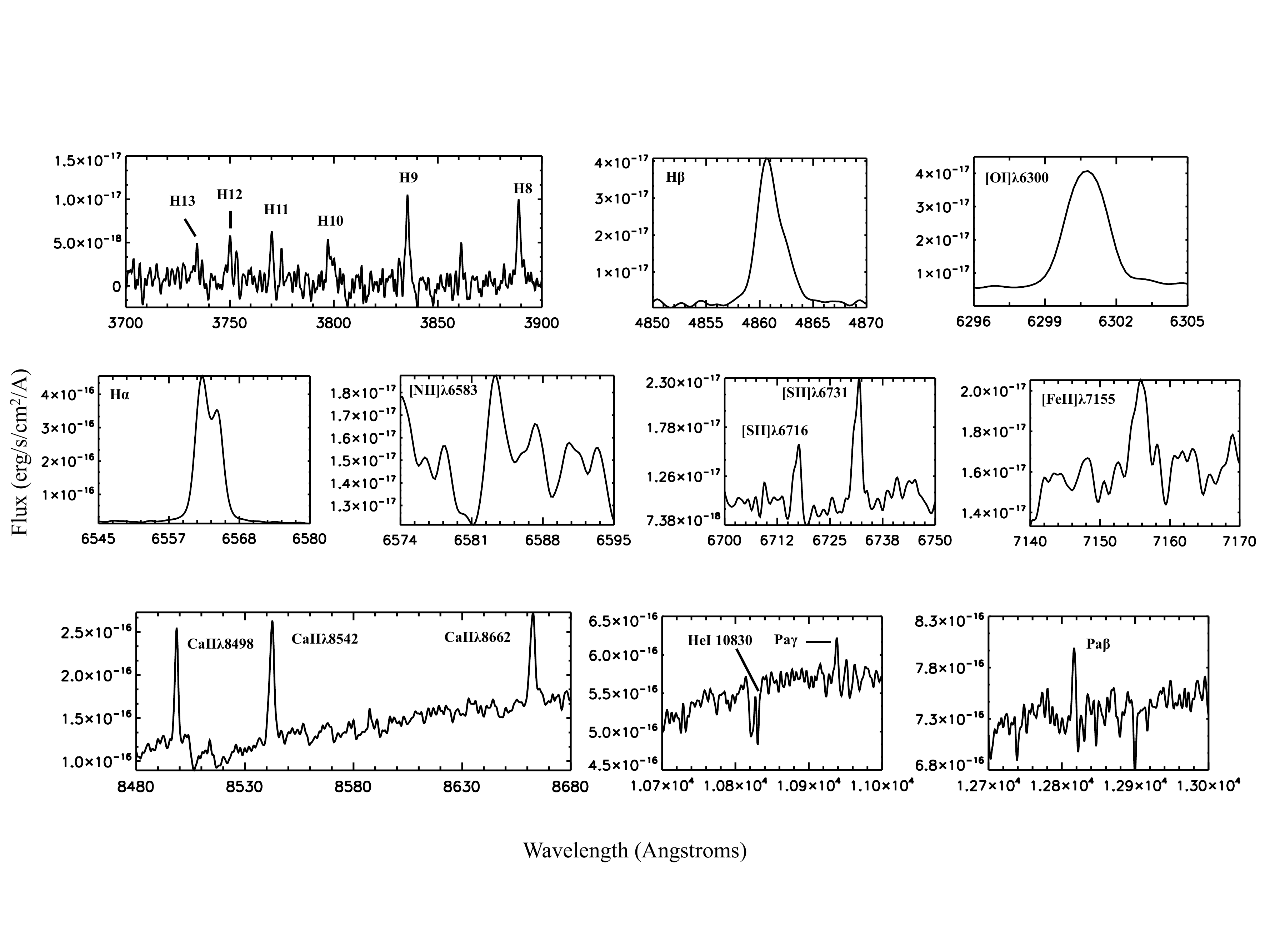}
     \caption{Spectrum as shown in Figure 1 with a zoom on key accretion and outflow tracers.}
  \label{nb_lines}     
\end{figure*}

\subsection{Data analysis}

The \xsh data reduction was performed independently for each arm using 
the X-Shooter pipeline version 1.3.7. The pipeline provides 2-dimensional, bias-subtracted, flat-field corrected, order-merged, background-subtracted 
and wavelength-calibrated spectra, but special attention was given to the 
sky subtraction. Sky regions free of nebular emission were selected along the
slit to derive a sky spectrum which was then subtracted to the 1-dimensional 
spectrum. Flux calibration was achieved within the pipeline using a response function derived from the spectra of flux-calibrated standard stars observed on the same night and of \iso. Following the independent flux calibration of the \xsh arms,  the internal
consistency of the calibration was checked by plotting together the three spectra extracted
from the source position and visually examining the superposition of overlapping
spectral regions at the edge of each arm. The UVB and VIS arms were found to be very well
aligned, while the NIR arm presented a shift with respect to the
VIS spectrum of $\sim$ 25$\%$ lower. This was corrected for by scaling the NIR spectrum to the VIS
continuum level. Finally, the correction for the contribution of telluric bands was done using the telluric standards observed with the same instrumental set-up, and close in airmass to \iso. More details on data reduction, as well as on flux calibration and correction for telluric bands can be found in \cite{Alcala14}.


The absolute velocity calibration provided by the \xsh pipeline was checked using the OH telluric lines distributed across the wavelength range of X-Shooter. We found an average difference of $\sim$ 9~\km\ between the calibration performed by the pipeline and that done using the telluric lines. In Figure \ref{shift} the measured velocities of the telluric lines over the wavelength range in which the outflow is detected is shown. The spectra were corrected for this velocity difference. All velocities given in this paper are also corrected for the stellar rest velocity of \iso\ measured with respect to the local standard of rest (LSR). In \cite{Whelan09b} the average velocity of YSOs in Cha~I of 2~\km\ was adopted for the stellar rest velocity of \iso. \cite{Joergens12} measured the rest velocity of \iso\ to be $\sim$6.6~\km, from the Li~I absorption feature at $\sim$ 6708~\AA. Here we adopt the value to be $\sim$ 6.4~\km\ as measured from the same Li~I feature.
For the position-velocity (PV) diagrams presented here, both the continuum emission and any sky lines were removed. This was done using the {\it continuum} routine within the Image Reduction and Analysis Facility (IRAF)\footnote{IRAF is distributed by the National Optical Astronomy Observatory, which is operated by the Association of the Universities for Research in astronomy. inc. (AURA) under cooperative agreement with the National Science Fundation }. Each continuum row or background column was plotted separately, the extent of the continuum or sky line fitted and then the fit was subtracted. For Figures 1, 4 and 5 Gaussian smoothing was used to increase the signal to noise ratio (SNR). This is commonly done in studies of BD outflows \citep{Whelan07, Whelan09b}.

\begin{figure*}
\centering
   \includegraphics[width=14cm, trim= 0cm 2.5cm 0cm 1cm]{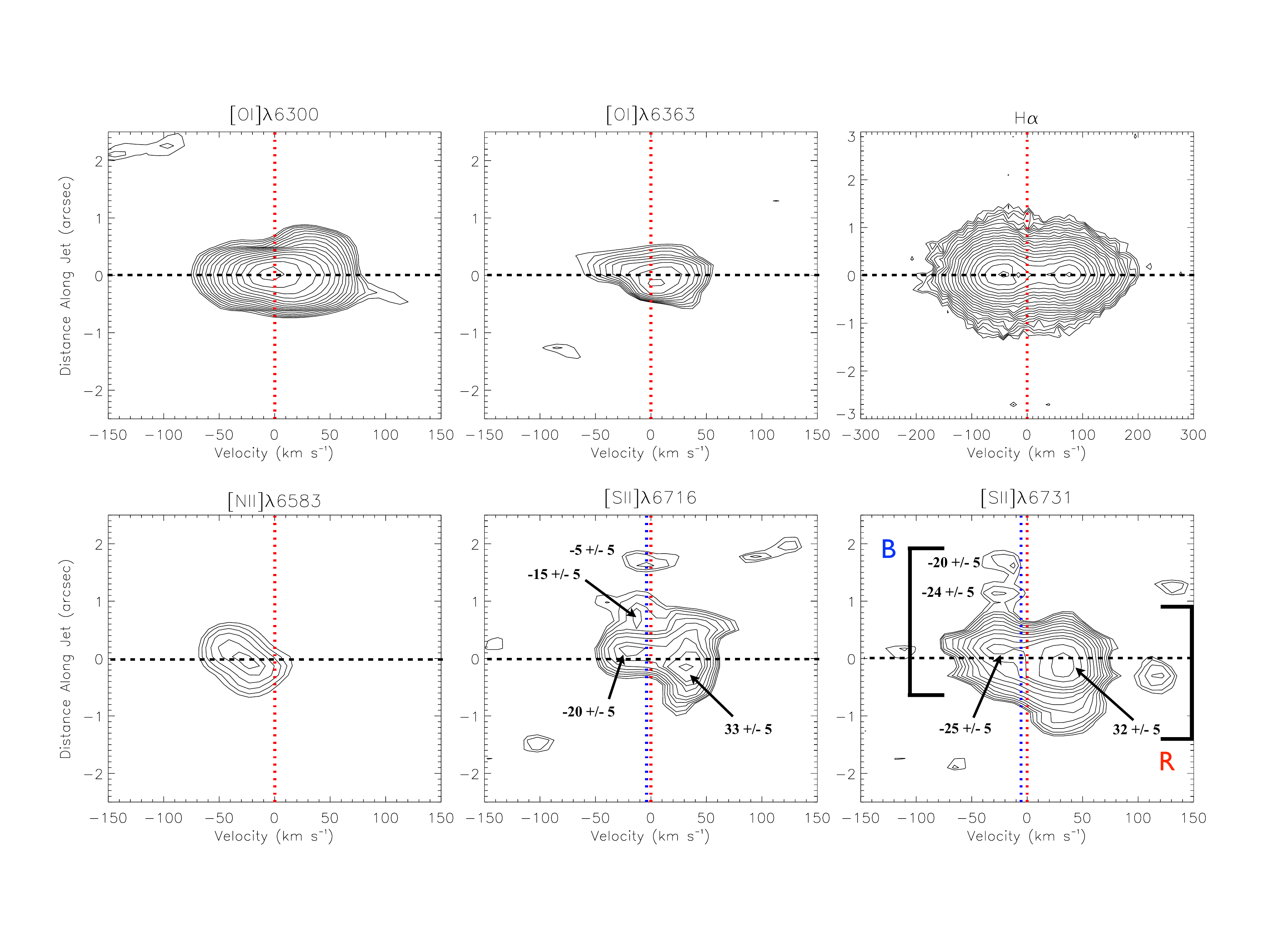}
     \caption{Contour plots for the main emission lines. Velocities are with respect to the systemic velocity of the source. The zero spatial position is the centre of the continuum emission, measured using Gaussian fitting, and is taken to be the source position. For the $[\ion{O}{i}]\, \lambda \lambda 6300,6363$ and H$\alpha$ lines, contours start at 3$\sigma$ and for the $[\ion{N}{ii}]\, \lambda 6583$ and $[\ion{S}{ii}]\, \lambda \lambda 6716,6731$ lines they start at 2$\sigma$. The 1$\sigma$ value is 1 $\times$ 10$^{-18}$ erg/s/cm$^{2}$. This is done to highlight the presence of faint extended blue-shifted [SII] emission and the non-detection of even fainter red-shifted [NII]$\lambda$6583 emission. In all cases, contours increase with a log scale ($\sqrt{1.5}$ for the [NII] and [SII] lines). The blue dashed lines mark the velocity of the faint [SII]$\lambda$6716, 6731 background lines. }
  \label{contour}     
\end{figure*}

\section{Results}

\subsection{Fundamental parameters of \iso}

The spectral type M6.5 ($\pm$0.5) we assign to \iso\ was derived as 
the average spectral type resulting from the various indices calculated 
from the VIS spectrum and following \citet{riddick07}. It is in agreement 
with determinations in the literature (see Section~2.1), and according 
to the  scale  by \cite{Luhman03} the corresponding temperature 
is $T_{\rm eff}=$2940\,K, is also in agreement with previous 
determinations. In \citet{Luhman07} an extinction of $A_{\rm J}=0.68$\,mag is derived,
meaning $A_{\rm V}=2.1$\,mag. Here the extinction law by \cite{Wein01} for R$_{V}$=5.5 was used. This law covers a wide range in wavelength, from the UV to the mid-IR. We estimated extinction in the same way as 
in our previous investigations \citep{Stelzer13, Alcala14, Whelan14}, 
that is, by finding the best match of artificially reddened X-shooter 
spectra of zero-extinction Class~III templates, with the spectrum of \iso. 
We derived $A_{\rm V}=2.5\pm0.3$\,mag, in fairly good agreement with the value 
reported by \cite{Luhman03}. 

Both mass and radius are crucial physical parameters for the estimates
of the mass accretion rate (see Section~4.1). The values derived by
\citet{Muz05} by comparison with the evolutionary models by
\citet{baraffe98} and \citet{chabrier00} are consistent with an object 
at the hydrogen burning limit. Another check for the temperature and radius can be done by determining 
the surface gravity, $\log{g}$, using spectral diagnostics and synthetic 
spectra. In order to estimate the surface gravity, $\log{g}$, and as a
by-product the projected rotational velocity, $v\sin{i}$, we used 
the gravity- and temperature-sensitive absorption doublets of $\ion{Na}{i}$ 
at $\lambda\lambda$ 8183.3, 8194.8\,nm and of $\ion{K}{i}$ at 
$\lambda\lambda$ 7664.8, 7698.9\,nm, by comparing the X-Shooter 
data to synthetic spectra. The BT-Settl model spectra of \citet{allard10} 
were used for a range of $T_{\rm eff}$ around the expected value and 
a range of $\log{g}$ values. 
The synthetic spectra were binned to the same spectral resolution as the X-Shooter data, and 
rotationally broadened in steps of $v\sin{i}$ of 5\,km\,s$^{-1}$ in the 
range from 10 to 35\,km\,s$^{-1}$. As an example, the synthetic spectra for 
$T_{\rm eff}$=2950\,K with two different values of $\log{g}$ are overlaid in Fig.~\ref{logg-vsini} 
on the X-shooter spectrum of \iso\ in the region of the $\ion{Na}{i}$ doublet. 
The model spectra have been rotationally broadened to $v\sin{i}=$20\,km\,s$^{-1}$, 
the rotation rate at which the models best fit the observed spectrum. 
The width of the observed lines is in agreement with a $\log{g}$ between 
3.5 and 4.0. Thus, the parameters for \iso,  $T_{\rm eff}=$2950\,K,  
$\log{g}=$3.7($\pm$0.3) and $v\sin{i}=$20($\pm$5)\,km\,s$^{-1}$ provide
the best fit to the data, with the uncertainties being due to the adopted 
steps in both $\log{g}$ and $v\sin{i}$ in the grid of synthetic spectra.
The gravity derived from the spectrum is in perfect agreement with the value 
calculated from the mass and radius as derived by \citet{Muz05} from evolutionary models. {\bf As a consequence, we confirm the results
on luminosity and age reported in \citet{Muz05} and \citet{Luhman07} (also see Section 2.1.)}

\subsection{Line identification}
The full UVB, VIS and NIR spectra of \iso\ are presented in Figure \ref{full}. By placing the \xsh slit along the outflow PA and using the nodding mode the outflow emission is traced to 8~\arcsec\ on either side of the driving source. The plotted 1D spectra were extracted by summing over the whole spatial extent of the 2D spectra and thus all the outflow emission is included. The only resolved emission along the slit is in the $[\ion{S}{ii}]\, \lambda \lambda 6716,6731$ lines and therefore there is no contribution from extended emission in any of the other lines. The identified lines are marked in Figure \ref{full}. Emission lines were identified using the Atomic Line List database
(http://www.pa.uky.edu/$\sim$peter/atomic/). For the identification, we considered
a wavelength uncertainty of about 0.5 \AA. Nebular lines from abundant
species having excitation energies below 40,000 cm$^{-1}$ were searched for. Line fluxes were computed through the Gaussian fitting of the line profiles after subtraction of the local continuum. 
Gaussian fitting was done using the IRAF task {\it splot}. Absolute flux errors
were computed from the root mean square (r.m.s) noise (measured in a portion of the spectrum adjacent
to the line) multiplied by the spectral resolution element at the considered wavelength.  The line fluxes are not corrected for extinction and a discussion on estimating the on-source extinction of \iso\ is given in Section 3.1. The spatial range over which the flux of the $[\ion{S}{ii}]\, \lambda \lambda 6716,6731$ was measured is marked in Figure 5. The same range is used for the $[\ion{N}{ii}]\, \lambda 6583$ emission and only blue-shifted emission is detected. In the case of the $[\ion{O}{i}]\, \lambda 6300$ line it is assumed that total flux is divided evenly between the two outflow lobes. For the other lines listed in Table 1 fluxes are measured from the 1D spectrum. 

In Figure \ref{nb_lines} line profiles of some of the most interesting lines are plotted. Firstly note that the Balmer lines from H13 to H$\alpha$ are detected. In Section 4.2 these lines are used to compute the Balmer decrements with respect to the H$\beta$ line. 
The H$\alpha$ line has a shape typical of an accretion dominated H$\alpha$ line region. While the outflow is not resolved in velocity in the $[\ion{O}{i}]\, \lambda 6300$ line the $[\ion{S}{ii}]\, \lambda \lambda 6716,6731$ line regions are double peaked (especially clear in the unsmoothed data) with the red-shifted peak being brighter. Thus we can extract velocity information from these lines and the \xsh line profile compares well to what was reported in \cite{Whelan09b}, in that the line is double peaked and the red-shifted peak is brighter. We discuss the $[\ion{S}{ii}]\, \lambda \lambda 6716,6731$ line regions further in Section 3.3. The CaII triplet, a further strong indicator of accretion is also detected. The HeI~1.083~$\mu$m line is seen in blue-shifted absorption. Finally the Pa$\gamma$ and Pa$\beta$ lines while faint are detected and are used in Section 4.1 to estimate $\dot{M}_{acc}$.



\subsection{Kinematics and morphology of the \iso\ jet} In \cite{Whelan09b} it was demonstrated using SA that the $[\ion{O}{i}]\, \lambda \lambda 6300,6363$ and $[\ion{S}{ii}]\, \lambda \lambda 6716,6731$
lines were formed in the outflow of \iso. As the spectra presented here were taken with the slit aligned parallel to the estimated outflow PA, it is now possible to investigate the kinematics and morphology of this outflow along the jet axis. 
In Figure \ref{contour} PV diagrams of the brightest outflow lines plus H$\alpha$ are presented. {\bf The H$\alpha$ line is the only line included in this figure that traces both outflow accretion as the FELs are quenched at the densities of the accretion shocks \citep{Hirth97}.} Our discussion of the kinematics and morphology of the outflow is based on this Figure. Firstly, from Figure \ref{contour} it is clear that the outflow is spatially resolved in the $[\ion{S}{ii}]\, \lambda \lambda 6716,6731$ lines. This confirms that the estimate of the outflow PA derived by \cite{Whelan09b} was accurate. The blue-shifted lobe extends to $\sim$ 1.6~\arcsec\ and the red-shifted lobe to $\sim$ 1~\arcsec. As described in Section 2 the velocity is measured with respect to the stellar rest velocity of \iso. In Figure \ref{contour} the velocities for the different components seen in the outflow are given. The red-shifted flow is marginally faster than the blue-shifted flow. Thus the velocity asymmetry reported in \cite{Whelan09b} and \cite{Joergens12} is not as pronounced in this dataset. For example, \cite{Whelan09b} gave the velocities of the blue and red-shifted lobes at -20~\km\ and -40~\km\ respectively. Therefore, in that study the ratio of velocity between the red and blue lobes is 2.0 as compared to the value of 1.4 found here. The difference between these two sets of velocities is due to the different values for the stellar rest velocity adopted (see Section 2) and also likely due to the fact that in \cite{Whelan09b} we were not observing along the outflow. Overall the radial velocities measured here can be considered to be more accurate.

As well as the small velocity asymmetry between the two lobes there is also a clear morphological asymmetry. The blue-shifted lobe has three separate knots while only one knot is resolved in the red-shifted lobe. Although the blue-shifted emission at $\sim$ 1~\arcsec\ and $\sim$ 1.6~\arcsec\ is faint we are satisfied that the fact that it is seen in both [SII] lines rules out the possibility that this is noise. 
We also consider the fact that the extended emission is due to an imperfect subtraction of faint background $[\ion{S}{ii}]\, \lambda \lambda 6716,6731$ emission. These background lines are a signature of nebular emission surrounding \iso\ and the velocity at which we see these lines is marked in Figure \ref{contour} by the blue dashed lines. For the $[\ion{S}{ii}]\, \lambda 6716$ and $[\ion{S}{ii}]\, \lambda 6731$ knots at $\sim$ 1~\arcsec\ there is enough of a separation in velocity from the background lines for us to conclude that these knots are real. Also the same argument applies for the $[\ion{S}{ii}]\, \lambda 6731$ knot at $\sim$ 1.6~\arcsec. The  $[\ion{S}{ii}]\, \lambda 6716$ emission at 1.6~\arcsec\ does lie at the velocity of the background line. Thus, we cannot categorically say without higher SNR data, that this is a knot in the outflow. However, the fact that its position coincides with a $[\ion{S}{ii}]\, \lambda 6731$ knot and the fact that we see no such extended emission in the other jet lines after sky subtraction makes it likely that it is a knot. By comparing the radial velocities of three [SII] knots in the blue-shifted outflow, it is seen that there is a small decrease in radial velocity along the blue-shifted flow.  A decrease in velocity with distance is often seen in jets from YSOs \citep{Davis03}. This can be explained by numerical simulations which show that speed of the shock front is lower than the initial velocity of ejection, due to the interaction with the ambient medium which decelerates the outflow \citep{Bonito04, Bonito07}. 

\begin{figure}[h!]
\centering
   \includegraphics[width=7cm, trim= 1.5cm 3.2cm 1.5cm 4cm]{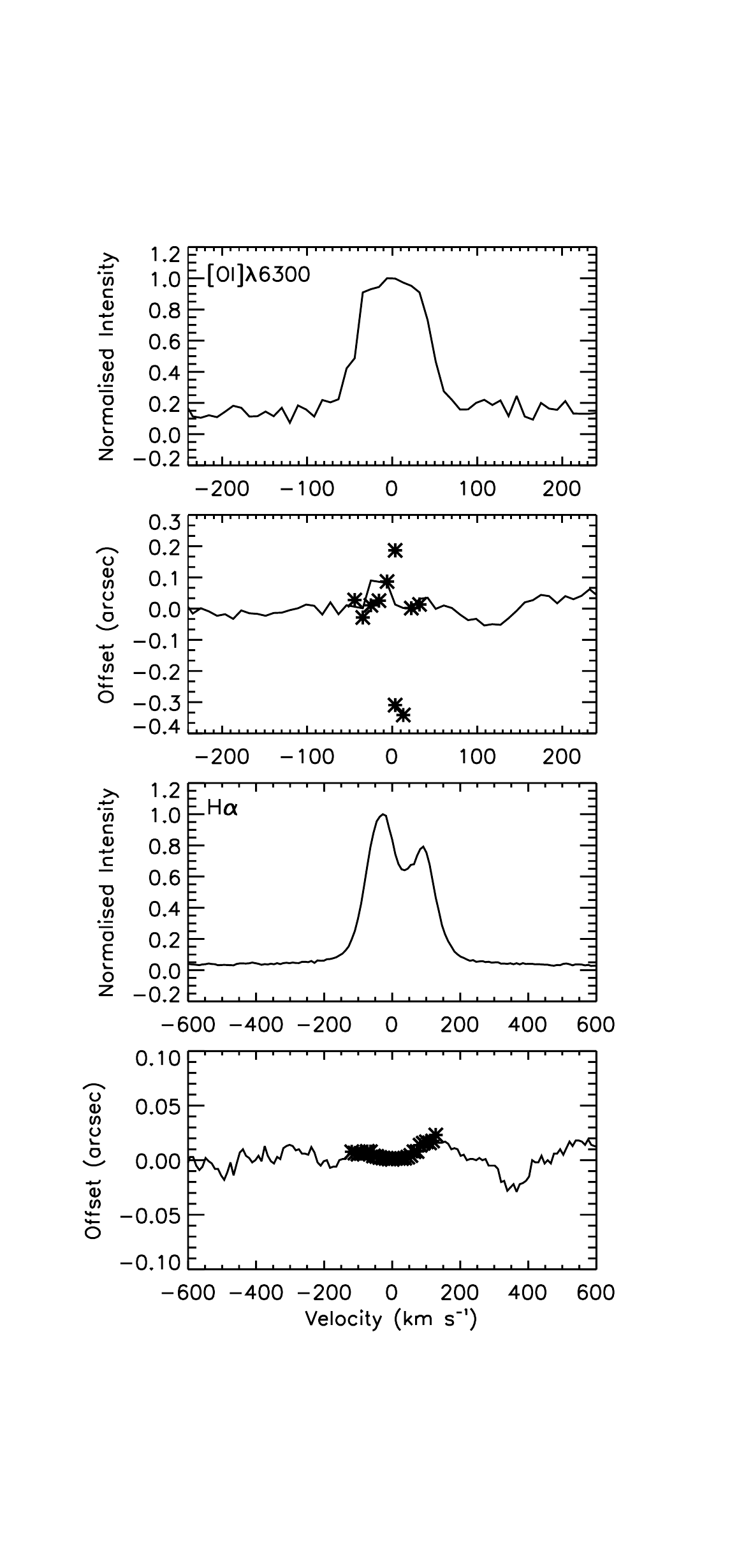}
     \caption{Spectro-astrometric analysis of the $[\ion{O}{i}]\, \lambda 6300$ and \Ha\ lines. Offsets are measured using Gaussian fitting. The black asteriks are the offsets measured after continuum subtraction. The accuracy in the measurement of the offsets is $\sim$ 20~mas. This analysis confims that the $[\ion{O}{i}]\, \lambda 6300$ line is tracing the outflow while no outflow component is detected in the \Ha\ line.} 
     \label{SA}    
\end{figure}

From Figure \ref{contour} it can also be seen that the outflow is not spatially resolved in the $[\ion{O}{i}]\, \lambda \lambda 6300,6363$ lines. This is not surprising as the critical density of the [OI] lines is two orders of magnitude higher than the critical density of the [SII] lines. Therefore [OI]  forms closer to the driving source of the outflow than the [SII] which often traces more extended emission \citep{Hartigan95}. This difference between the spatial extent of the $[\ion{O}{i}]\, \lambda \lambda 6300,6363$ and $[\ion{S}{ii}]\, \lambda \lambda 6716,6731$ lines was also noted by \cite{Whelan09b}. \cite{Whelan09b} measured offsets of  $\sim$ 200~mas in the [SII] lines while the [OI] were only extended to $\sim$ 50~mas. In Figure \ref{SA} the results of applying SA to the \xsh $[\ion{O}{i}]\, \lambda 6300$ line region are presented. This analysis confirms that the [OI] emission detected in the \xsh spectrum is formed in the \iso\ outflow, and the offsets are larger than those measured in \cite{Whelan09b} as expected for a spectrum taken along the outflow PA.  In Figure \ref{SA} the spectro-astrometric analysis of the H$\alpha$ line is also shown. We detect no outflow component to the H$\alpha$ line.  

Finally, and very interestingly, it can be seen from Figure \ref{contour} that $[\ion{N}{ii}]\, \lambda 6583$ is only detected in the blue-shifted flow and that it traces slightly higher velocities than the $[\ion{S}{ii}]\, \lambda \lambda 6716,6731$ lines. The estimated flux of the $[\ion{N}{ii}]\, \lambda 6583$ emission is 1.2 $\pm$ 0.2 $\times$ 10$^{-17}$ erg/s/cm$^{2}$. The critical density of the blue-shifted $[\ion{N}{ii}]\, \lambda 6583$ emission is $\sim$ 6 $\times$ 10$^{4}$ cm$^{-3}$ at 10,000~K and it only traces the highest velocity emission in YSO jets \citep{Hirth97}. The non-detection of $[\ion{N}{ii}]\, \lambda 6583$ in the red-shifted flow implies that the ionisation fraction is smaller in the red-shifted lobe than in the blue-shifted lobe (see Table 2). {\bf There are many examples of CTT jets where $[\ion{N}{ii}]\, \lambda 6583$ is only weakly detected compared to other FELs and thus a low ionisation is inferred. For example \citep{Coffey08} discuss the weak $[\ion{N}{ii}]\, \lambda 6583$ emission from and thus low ionisation of the RW Aur red-shifted jet. Furthermore, better angular resolution observations are needed before one can conclude if partial obscuration by the \iso\  disk could also contribute to the lack of $[\ion{N}{ii}]\, \lambda 6583$ emission from the \iso\ red-shifted jet. }



\section{Discussion}

\subsection{Estimating $\dot{M}_{out}$/$\dot{M}_{acc}$}

The mass accretion rate was calculated from the following equation 
\begin{equation}
\dot{M}_{acc} = 1.25 (L_{acc} R_{*}) / (G M_{*})
 \label{Macc}
\end{equation}
\noindent where L$_{acc}$ is the accretion luminosity and R$_{*}$ and M$_{*}$ are the stellar radius and mass \citep{Gullbring98, Hartmann98}. L$_{acc}$ is derived from the luminosity of 13 accretion tracers (L$_{line}$; Figure 8) using the relationships published in \cite{Alcala14}. L$_{line}$ was calculated from the extinction corrected fluxes (see section 3.1) of the accretion indicators and assuming a  distance of 140~pc. {\bf Due to the moderate inclination of the disk (given in Section 2.1) L$_{line}$ did not need to be corrected for obscuration by the disk as in \cite{Whelan14}.} Also included is the literature value taken from \citep{Muz05}. We also consider the effect of variable accretion by comparing log($\dot{M}_{acc}$) derived here from the H$\alpha$ line to the value derived from the VLT / UVES spectra presented in \cite{Whelan09b}. It can be seen from Figure 7 that these values are compatible within the errors. It is found that the average value for log($\dot{M}_{acc}$) [\Msun/yr]= -10.6 $\pm$ 0.4 which is consistent within errors with the result of \citep{Muz05}.


In Table 2 the physical parameters of the \iso\ blue and red jets and the derived values of $\dot{M}_{out}$ are given. The physical parameters are the electron density ($n_{e}$), the electron temperature ($T_{e}$), and the ionisation fraction ($x_{e}$). The electron density is estimated from the ratio of the $[\ion{S}{ii}]\, \lambda \lambda 6716,6731$, $T_{e}$ from the ratio $[\ion{O}{i}]\, \lambda 6300$/ $[\ion{S}{ii}]\, \lambda 6731$ and x$_{e}$ from the ratio $[\ion{N}{ii}]\, \lambda 6583$ / $[\ion{O}{i}]\, \lambda 6300$ \citep{Whelan14}. In the case of the ionisation in the red-shifted jet an upper limit of 3 times the r.m.s noise was used for the flux of the $[\ion{N}{ii}]\, \lambda 6583$ line. The extent to which a YSO jet is extincted by the circumstellar material is an important parameter when calculating $\dot{M}_{out}$ and can be estimated using the [Fe II] jet lines at 1.27, 1.32, 1.64 $\mu$m \citep{Nisini05}. As these lines are not detected in the spectrum of \iso\ it is not possible to know the level to which the \iso\ jet is extincted by any circumstellar material. Therefore, to test the dependence of the results on the jet extinction, the physical parameters and $\dot{M}_{out}$ were calculated for a range of values of A$_{v}$ as shown in Table 2. Increasing the extinction and thus the line flux does not substantially effect the values of n$_{e}$, $T_{e}$ and x$_{e}$. It mostly impacts L$_{SII}$ and thus $\dot{M}_{out}$. 

The measurements of $\dot{M}_{out}$ are based on the following equations and further information can be found in \cite{Whelan14}.
\begin{equation}
\dot{M}_{out} = \mu\,m_{H}\,(n_{H}\,V)\,V_{tan}/l_{t}
\end{equation}
with 
\begin{equation}
n_H\,V = L_{SII}\,\left(h\,\nu\,A_{i}\,f_{i}\,\frac{X^i}{X}\,\frac{X}{H}\right)^{-1}
\end{equation}
Here $\mu=1.24 $ is the mean atomic weight, $m_{H}$ the proton mass,  
$V$ the volume effectively filled by the emitting gas,
$V_{tan}$ and $l_{t}$ the tangential velocity and length of the knot, 
$A_{i}$ and $f_{i}$ the radiative rate 
and upper level population relative to the considered transition and finally
$\frac{X^i}{X}$ and $\frac{X}{H}$ are x$_{e}$ and the relative 
abundance of the considered species. The ionisation fraction and the upper level population are estimated from the physical parameters given in Table 2. The tangential velocities of the blue and red jets are taken at 25~\km\ and 32~\km\ respectively, and l$_{t}$ = 1~\arcsec. We do not include the outer blue-shifted knot in our calculation as we should only include knots which are detected in either [OI]$\lambda$6300 or [NII]$\lambda$6583. L$_{SII}$ is the luminosity of the [SII]$\lambda$6731 line and it is assumed in this calculation that all the sulphur is single ionised.


Overall it is concluded from the results presented in Table 2 that $\dot{M}_{out}$/$\dot{M}_{acc}$ (two-sided) for \iso\ is consistent with magneto-centrifugal jet launching models and studies of CTTSs. This conclusion applies for all chosen values of A$_{v}$. In  \cite{Whelan09b} $\dot{M}_{out}$ was measured at 1.8 $\times$ 10$^{-10}$ \Msun yr$^{-1}$ and 3.1 $\times$ 10$^{-10}$ \Msun yr$^{-1}$ for the blue-shifted and red-shifted jets respectively and thus $\dot{M}_{out}$/$\dot{M}_{acc}$ using the value of $\dot{M}_{acc}$ published by \cite{Muz05} was $>$ 1. 
In \cite{Whelan09b} the extinction of the source and the jet was assumed to be the same. The main reason for the difference between the lower and thus more plausible estimate of $\dot{M}_{out}$/$\dot{M}_{acc}$ made here, and the value published in \cite{Whelan09b} is that in \cite{Whelan09b} the method used assumed a value for the critical density (n$_{cr}$) which was highly uncertain. Here an estimate of n$_{cr}$ is not required. Estimates of $\dot{M}_{acc}$ are more accurate due to the use of several accretion tracers. Previous to this work it has only been derived from the H$\alpha$ line. However, this change in the derived value of $\dot{M}_{acc}$ does not have a significant effect on the final values of $\dot{M}_{out}$/$\dot{M}_{acc}$. Similarily extinction does not have a large effect on our conclusion in this case. As for all values of A$_{v}$ up to the value measured for the driving source $\dot{M}_{out}$/$\dot{M}_{acc}$ remains within the limits of predictions by leading models.

\begin{figure}
\centering
   \includegraphics[width=9.5cm, trim= 1cm 1cm 0cm 3.1cm]{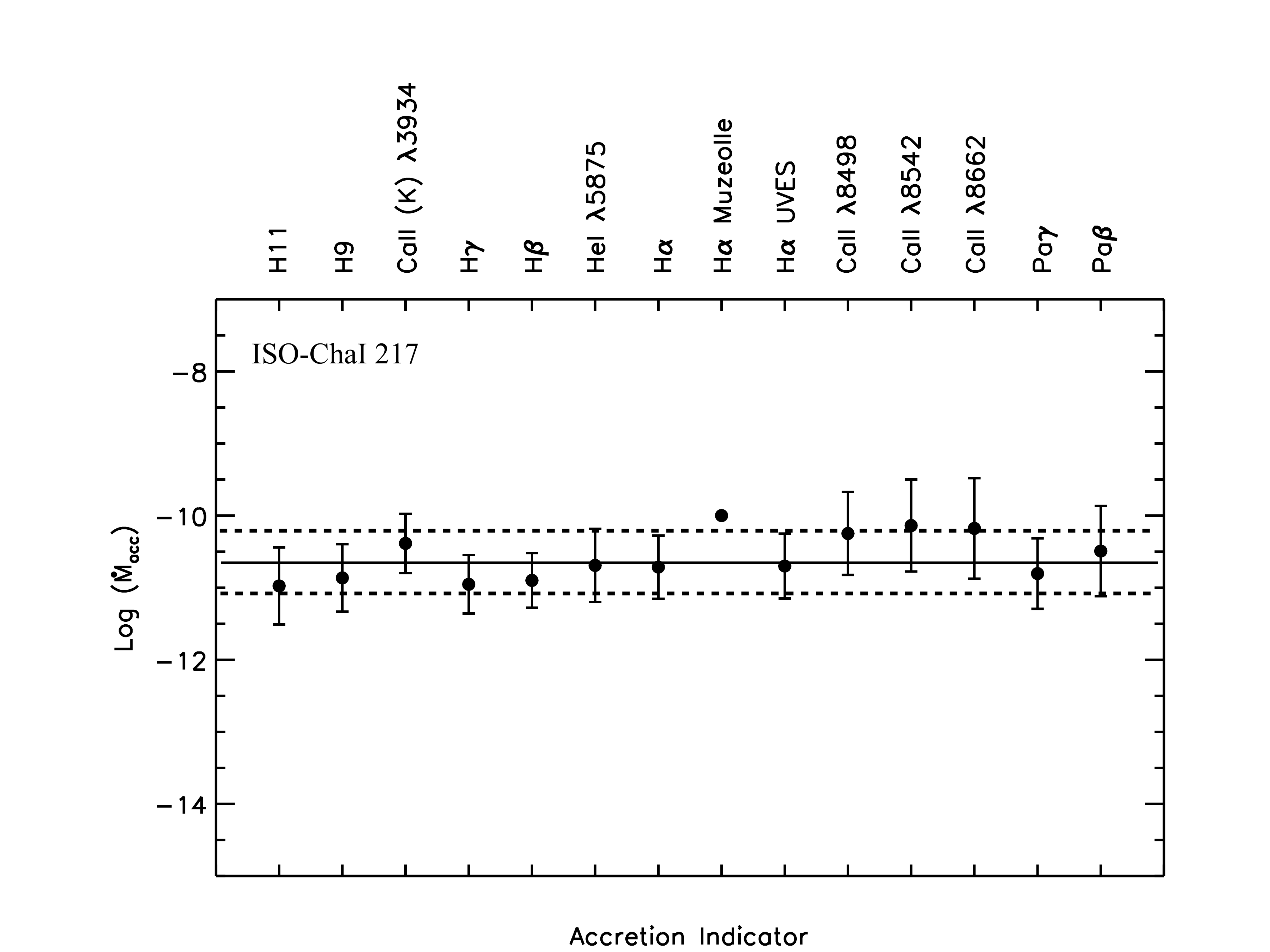}
     \caption{Mass accretion rate estimated from various accretion tracers. The solid line gives the mean value of log ($\dot{M}_{acc}$) and the dashed line is the $\pm$ 1$\sigma$ uncertainty. The errors on the individual measurements mainly come from the error in the line fluxes and in the value of the on-source extinction. The value of $\dot{M}_{acc}$ calculated by \cite{Muz05} from modelling of the H$\alpha$ line and from the UVES spectra presented by \cite{Whelan09b} are included here for comparison. No error estimates were given in \cite{Muz05}.}
  \label{acc}     
\end{figure}



\begin{table*}
\begin{center}
\begin{tabular}{llll} 
\hline\hline
A$_{v}$ (mag)  &0.0  &1.0 &2.5 \\
\hline
 n$_{e}$ Blue (cm$^{-3}$)                        &4610                     &4700                   &4920                                        
\\
 n$_{e}$ Red (cm$^{-3}$                         &5490                     &5630                    &5750                                   
\\
$T_{e}$ Blue (10$^{4}$ K)                       &2.15                             &2.24                            &2.34                                     
\\
$T_{e}$ Red (10$^{4}$ K)                       &1.63                             &1.71                             &1.81 
\\
x$_{e}$ Blue                                             &0.078                             &0.063                             &0.048 
\\
x$_{e}$ Red                                             &0.045                             &0.040                            &0.034                                    
\\
n$_{H}$ Blue  (10$^{4}$ cm$^{-3}$)        &6.0 $\pm$ 0.8                         &7.5 $\pm$ 1.0                         &10.3 $\pm$ 1.4
\\
n$_{H}$ Red   (10$^{4}$ cm$^{-3}$)        &12.2 $\pm$ 4.4                             &14.0 $\pm$ 5.0                           &17.0 $\pm$ 6.2
\\
\hline
\\
Method B & & &
\\
L$_{SII}$ Blue (10$^{-8}$ \Lsun)   &1.1 $\pm$ 0.3  &2.3 $\pm$ 0.4 &5.6 $\pm$ 1.0 
\\
L$_{SII}$ Red (10$^{-8}$ \Lsun)   &1.4 $\pm$ 0.3  &2.8 $\pm$ 0.5 &6.9 $\pm$ 1.2 
\\
$\dot{M}_{out}$ (10$^{-12}$  \Msun yr$^{-1}$) Blue &0.7 $\pm$ 0.2 &1.4 $\pm$ 0.3 &3.3 $\pm$ 0.7 
\\
$\dot{M}_{out}$ (10$^{-12}$  \Msun yr$^{-1}$) Red &1.2 $\pm$ 0.5 &2.3 $\pm$  0.9 &5.3 $\pm$ 2.1
\\
($\dot{M}_{out}$ Blue + $\dot{M}_{out}$ Red) / $\dot{M}_{acc}$ &0.05 (+0.07)(-0.02)  &0.09 (+0.14)(-0.04)  &0.20 (+0.30)(-0.09) 
\\
\hline
\end{tabular}
\caption{The jet physical parameters and $\dot{M}_{out}$ for the \iso\ blue and red jets. A$_{v}$ here refers to the extinction of the jet and the calculations are made for three values of  A$_{v}$ to investigate the dependence on the jet extinction. The mean value of $\dot{M}_{acc}$ ($\dot{M}_{acc}$ mean = 4 $\times$ 10$^{-11}$ \Msun yr$^{-1}$) is used to calculate $\dot{M}_{out}$/$\dot{M}_{acc}$ and $\dot{M}_{acc}$ is derived from the fluxes of the accretion tracers listed in Figure 8 corrected for an on-source extinction  2.5 $\pm$ 0.3~mag.}  
\end{center}
\end{table*}


\subsection{Investigating the origin of the HI emission}
\begin{figure}[h!]
\centering
   \includegraphics[width=9cm, trim= 0cm 0cm 0cm 2cm, clip=true]{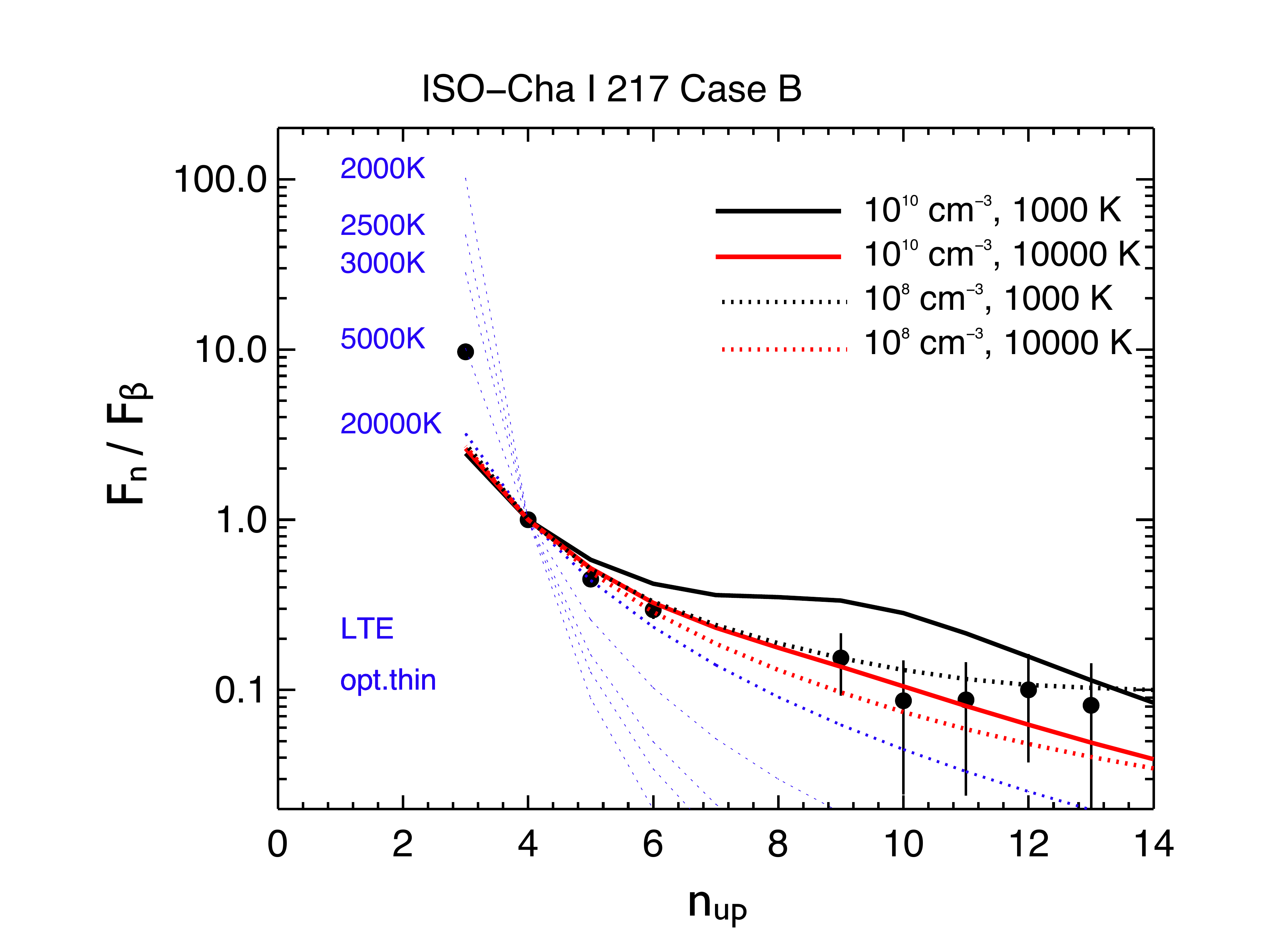}
     \caption{Balmer decrements. Results seem to be consistent with an accretion scenario, similar to \para. }
  \label{balmer}     
\end{figure}

Several HI emission lines are detected in the spectrum of \iso\ (see Figure \ref{nb_lines}). To investigate the origin of this emission the Balmer decrements were computed with respect to H$\beta$ line and are plotted as a function of  upper quantum number (n$_{up}$) in Figure \ref{balmer}. While the spectral range of X-Shooter covers most of the Balmer lines any lines with n$_{up}$ $>$ 13 were found to be too noisy for inclusion in our analysis. Additionally, at the intermediate spectral resolution of X-Shooter the H7 and H8 lines are blended with other lines and are therefore also not included. To constrain the physical conditions in the emitting gas the decrements were firstly compared to standard Case B predictions calculated for a range of temperature and density. The Case B curves were derived using the calculations of \cite{Hummer87} and using the Fortran program and data files provided by \cite{Storey95}. These models assume that all lines are optically thin. Secondly, the decrements were also compared to optically thick and thin local thermodynamic equilibrium cases (LTE) ratios, calculated over a temperature range of 2000 K to 20000 K. The optically thick case is not shown in Figure \ref{balmer} as it was not a good fit to the results. The emission is best fit with the Case B model and $T_{e}$ =10,000~K, n$_{e}$ = 10$^{10}$~cm$^{-3}$. This suggests formation in an accretion flow \citep{Martin96}. This is consistent with the spectro-astrometric analysis of the H$\alpha$ line (Figure 6) in which no outflow component is detected. It is likely that the outflow can be traced by the H$\alpha$ line but the emission is many times fainter than the accretion component. Note that no Brackett lines and only two Paschen lines are detected.

\section{Summary}
Here we present a follow-up X-Shooter study of the BD candidate \iso\ and its outflow. Previous to this work \iso\ was found to have a bipolar outflow which exhibited various asymmetries and $\dot{M}_{out}$/$\dot{M}_{acc}$ had been estimated to be $>$ 1. The advantage of using X-Shooter is that the source and outflow can be probed at wavelengths previously not investigated and that $\dot{M}_{acc}$ and $\dot{M}_{out}$/$\dot{M}_{acc}$ can be more accurately estimated. An additional advantage of this study is that data were collected with the slit aligned with the derived outflow PA thus allowing the technique for estimating the PA of outflows using SA to be tested \citep{Whelan09b} and, the morphology and kinematics of the \iso\ jet studied with distance along the jet. The results can be summarised as follows. 

\begin{itemize}

\item{The outflow is spatially resolved in the [SII]$\lambda\lambda$6716, 6731 lines and three separate knots at $\sim$ 0\farcs1, 0\farcs7 and 1\farcs6 are detected in the blue-shifted flow. Red-shifted emission extends to $\sim$ 1~\arcsec\ with an emission peak at 0\farcs2.}

\item{While the velocity asymmetry between the blue-shifted and red-shifted lobes of the outflow is not as pronounced as reported in \cite{Whelan09b} the red-shifted flow is still found to be faster than the blue-shifted flow. In \cite{Whelan09b} the factor by which it was faster was $\sim$ 2 while here it is $\sim$ 1.4. There are also morphological asymmetries in that three knots are detected in the blue-shifted jet while only one red-shifted knot is detected. These kinematical and morphological asymmetries can be explained by a pulsed jet model which describes how if different ejection rates are involved in the opposite sides of a bipolar jet, different knot velocities are expected \citep{Bonito2010a, Bonito2010b}. Also $\dot{M}_{out}$ is larger in the red-shifted jet than in the blue-shifted, even when we account for the uncertainty introduced by the use of an upper limit for the flux of the $[\ion{N}{ii}]\, \lambda 6583$ line in the red-shifted jet. This difference in $\dot{M}_{out}$ was also found in \cite{Whelan09b} and this sort of of asymmetry can be reproduced by the lastest models which account for jet asymmetries \citep{Fendt13}.}

\item{Using the improved method for estimate $\dot{M}_{out}$ we now place $\dot{M}_{out}$/$\dot{M}_{acc}$ (for the \iso\ jets combined) at  0.05 (+0.07)(-0.02). This value assumes that the outflow is not affected by extinction and unlike previous estimates this ratio is now in agreement with predictions of jet launching models  \citep{Ferreira06}. This agreement still persists even if we assume that the extinction affecting the outflow is comparable to the value of A$_{v}$ measured here for the source. This is significant as it is one of the few studies which show this ratio to be the same in BDs as in CTTS \citep{Stelzer13, Whelan14}.}

\item{Finally by analysing the Balmer decrements of the selection of Balmer lines found in the \iso\ spectrum it is concluded that the emission is best fit with a temperature of 10,000~K and a density of 10$^{10}~$cm$^{-3}$. This confirms that the bulk of the Balmer emission comes from the accretion flow in agreement with the spectro-astrometric analysis of the H$\alpha$ line.}

\end{itemize}

Overall this work adds to the similarities already observed between BD and CTT outflows. Now as well as having kinematical and morphological properties which are comparable to CTT jets $\dot{M}_{out}$/$\dot{M}_{acc}$ is also comparable in a few cases. Additionally, it demonstrates the usefulness of SA for determining basic information about a BD outflow such as the outflow PA, and the value of following up such spectro-astrometric investigations with high quality spectroscopic observations made parallel to the jet axis. The methods used here should be applied to a larger number of BD outflows to better constrain $\dot{M}_{out}$/$\dot{M}_{acc}$ at the lowest masses.

\acknowledgements{E.T. Whelan acknowledges financial support from the BMWi/DLR grant FKZ 50 OR 1309 and from the Deutsche Forschungsgemeinschaft through the Research Grant Wh 172/1-1. This work was financially supported by the PRIN INAF 2013 "Disks, jets and the dawn of planets". We thank G. Cupani, V. D'Odorico, P. Goldoni and A. Modigliani
 for their help with the X-Shooter pipeline. 
 are acknowledgement for the installation of the different  pipeline
 versions at Capodimonte.
 We also thank the ESO staff, in particular F. Patat for suggestions
 in OB preparation and C. Martayan for support during the observations. Finally we would also like to thank the anonymous referee for their comments.
}

{}


\end{document}